\documentclass[aps,eqsecnum,preprint,preprintnumbers,12pt,amsfonts]{revtex4}
\usepackage{epsfig,psfrag}

\usepackage{latexsym}
\usepackage{graphicx}
\usepackage{axodraw}

\newcommand{\be}{\begin{equation}}
\newcommand{\ee}{\end{equation}}
\newcommand{\ba}{\begin{eqnarray}}
\newcommand{\ea}{\end{eqnarray}}
\def\nn{\nonumber}
\def\half{\frac{1}{2}}
\def\Eins{\mathord{1\hskip -1.5pt
\vrule width .5pt height 7.75pt depth -.2pt \hskip -1.2pt
\vrule width 2.5pt height .3pt depth -.05pt \hskip 1.5pt}}

\def\olmass{\begin{picture}(10,10)(10,10)
\PhotonArc(15,10)(10,0,180)2 5
\Line(0,10)(30,10)
\end{picture}}

\def\tlscbf{\begin{picture}(10,10)(10,10)
\Photon(4,15)(26,15)2 4
\CArc(15,15)(10,0,180)
\CArc(15,15)(10,180,360)
\CArc(15,15)(12,0,180)
\CArc(15,15)(12,180,360)
\end{picture}}

\def\tlscbfph{\begin{picture}(10,10)(10,10)
\DashLine(4,15)(26,15){1}
\CArc(15,15)(10,0,180)
\CArc(15,15)(10,180,360)
\CArc(15,15)(12,0,180)
\CArc(15,15)(12,180,360)
\end{picture}}

\def\tlscbfphvv{\begin{picture}(10,10)(10,10)
\DashLine(4,15)(26,15){1}
\CArc(15,15)(10,0,180)
\CArc(15,15)(10,180,360)
\CArc(15,15)(12,0,180)
\CArc(15,15)(12,180,360)
\Vertex(15,15)2
\put(13,18){\tiny 2}
\end{picture}}

\def\tlscbfphvvv{\begin{picture}(10,10)(10,10)
\DashLine(4,15)(26,15){1}
\CArc(15,15)(10,0,180)
\CArc(15,15)(10,180,360)
\CArc(15,15)(12,0,180)
\CArc(15,15)(12,180,360)
\Vertex(15,15)2
\put(13,18){\tiny 3}
\end{picture}}

\def\tlscbfphvvvv{\begin{picture}(10,10)(10,10)
\DashLine(4,15)(26,15){1}
\CArc(15,15)(10,0,180)
\CArc(15,15)(10,180,360)
\CArc(15,15)(12,0,180)
\CArc(15,15)(12,180,360)
\Vertex(15,15)2
\put(13,18){\tiny 4}
\end{picture}}

\def\tlscbfphvnn{\begin{picture}(10,10)(10,10)
\DashLine(4,15)(26,15){1}
\CArc(15,15)(10,0,180)
\CArc(15,15)(10,180,360)
\CArc(15,15)(12,0,180)
\CArc(15,15)(12,180,360)
\Vertex(15,15)2
\put(10,18){\tiny n-1}
\end{picture}}

\def\tlscbfphvn{\begin{picture}(10,10)(10,10)
\DashLine(4,15)(26,15){1}
\CArc(15,15)(10,0,180)
\CArc(15,15)(10,180,360)
\CArc(15,15)(12,0,180)
\CArc(15,15)(12,180,360)
\Vertex(15,15)2
\put(13,18){\scriptsize n}
\end{picture}}

\def\tlscbfphvbn{\begin{picture}(10,10)(10,10)
\DashLine(4,15)(26,15){1}
\CArc(15,15)(10,0,180)
\CArc(15,15)(10,180,360)
\CArc(15,15)(12,0,180)
\CArc(15,15)(12,180,360)
\Vertex(15,15)2
\put(12,18){\tiny N}
\end{picture}}

\def\tlscbfphvnnn{\begin{picture}(10,10)(10,10)
\DashLine(4,15)(26,15){1}
\CArc(15,15)(10,0,180)
\CArc(15,15)(10,180,360)
\CArc(15,15)(12,0,180)
\CArc(15,15)(12,180,360)
\Vertex(15,15)2
\put(8,18){\tiny n+1}
\end{picture}}

\def\tlscfrph{\begin{picture}(10,10)(10,10)
\DashLine(5,15)(25,15){1}
\CArc(15,15)(10,0,180)
\CArc(15,15)(10,180,360)
\end{picture}}

\def\tlscfrphvv{\begin{picture}(10,10)(10,10)
\DashLine(5,15)(25,15){1}
\CArc(15,15)(10,0,180)
\CArc(15,15)(10,180,360)
\Vertex(15,15)2
\put(13,18){\tiny 2}
\end{picture}}

\def\tlscfrphvvv{\begin{picture}(10,10)(10,10)
\DashLine(5,15)(25,15){1}
\CArc(15,15)(10,0,180)
\CArc(15,15)(10,180,360)
\Vertex(15,15)2
\put(13,18){\tiny 3}
\end{picture}}

\def\tlscfrphvvvv{\begin{picture}(10,10)(10,10)
\DashLine(5,15)(25,15){1}
\CArc(15,15)(10,0,180)
\CArc(15,15)(10,180,360)
\Vertex(15,15)2
\put(13,18){\tiny 4}
\end{picture}}

\def\tlscfrphvn{\begin{picture}(10,10)(10,10)
\DashLine(5,15)(25,15){1}
\CArc(15,15)(10,0,180)
\CArc(15,15)(10,180,360)
\Vertex(15,15)2
\put(13,18){\tiny n}
\end{picture}}

\def\tlscfrphvnn{\begin{picture}(10,10)(10,10)
\DashLine(5,15)(25,15){1}
\CArc(15,15)(10,0,180)
\CArc(15,15)(10,180,360)
\Vertex(15,15)2
\put(10,18){\tiny n-1}
\end{picture}}

\def\tlscfrphvvvvggg{\begin{picture}(10,10)(10,10)
\DashLine(5,15)(25,15){1}
\CArc(15,15)(10,0,180)
\CArc(15,15)(10,180,360)
\Vertex(15,15)2
\Vertex(15,25)2
\put(13,29){\tiny 3}
\put(13,18){\tiny 4}
\end{picture}}

\def\tlscfr{\begin{picture}(10,10)(10,10)
\Photon(5,15)(25,15)2 4
\CArc(15,15)(10,0,180)
\CArc(15,15)(10,180,360)
\end{picture}}

\def\olscbf{\begin{picture}(10,10)(10,10)
\CArc(15,15)(10,0,180)
\CArc(15,15)(10,180,360)
\CArc(15,15)(12,0,180)
\CArc(15,15)(12,180,360)
\end{picture}}

\def\olscfr{\begin{picture}(10,10)(10,10)
\CArc(15,15)(10,0,180)
\CArc(15,15)(10,180,360)
\end{picture}}

\def\olscfrvdva{\begin{picture}(10,10)(10,10)
\CArc(15,15)(10,0,180)
\CArc(15,15)(10,180,360)
\Vertex(15,5)2
\put(13,-4){\scriptsize 2}
\end{picture}}

\def\olscbfvdva{\begin{picture}(10,10)(10,10)
\CArc(15,15)(10,0,180)
\CArc(15,15)(10,180,360)
\CArc(15,15)(12,0,180)
\CArc(15,15)(12,180,360)
\Vertex(15,5)2
\put(13,-4){\scriptsize 2}
\end{picture}}

\def\olscfrvst{\begin{picture}(10,10)(10,10)
\CArc(15,15)(10,0,180)
\CArc(15,15)(10,180,360)
\Vertex(15,5)2
\put(13,-4){\scriptsize 4}
\end{picture}}

\def\olscfrvn{\begin{picture}(10,10)(10,10)
\CArc(15,15)(10,0,180)
\CArc(15,15)(10,180,360)
\Vertex(15,5)2
\put(13,-4){\scriptsize n}
\end{picture}}

\begin{document}

\title{Two-Loop Vacuum Diagrams in Background Field and the Heisenberg-Euler Effective Action}

\author{Marek Kras\v nansk\'y\footnote{mkras@phys.uconn.edu}}

\affiliation{Department of Physics,
University of Connecticut,
Storrs, CT 06269-3046, USA}

\begin{abstract}
We show that in arbitrary even dimensions, the two-loop scalar QED
Heisenberg-Euler effective action can be reduced to simple one-loop
quantities, using just algebraic manipulations, when the constant
background field satisfies $F^2=-f^2 \Eins$, which in four
dimensions coincides with the condition for self-duality, or
definite helicity. This result relies on new recursion relations
between two-loop and one-loop diagrams, with background field
propagators. It also yields an explicit form of the renormalized
two-loop effective action in a general constant background field in
two dimensions.
\end{abstract}

\maketitle

\section{Introduction}

   Great progress has been made in developing techniques of calculating multi-loop Feynman diagrams
\cite{Smirnov, Steinhauser, Baikov, Bern, Tarasov, Kotikov, Laporta, Broadhurst, Vladimirov, Bender, Chetyrkin_Tkachov,
Anastasiou, Avdeev, Baikov2, Smirnov_Steinhauser, Schroder, Chetyrkin_Faisst_Sturm_Tentyukov}. The "integration by parts" method,
allowing to reduce higher order diagrams into a set of basic integrals, offers a powerful tool for calculating multi-loop
amplitudes. This method has been applied to massless \cite{Bender, Chetyrkin_Tkachov, Anastasiou} as well as massive propagators
\cite{Avdeev, Baikov2, Smirnov_Steinhauser, Schroder, Chetyrkin_Faisst_Sturm_Tentyukov}. Recently such algebraic methods proved
to be very useful also for diagrams containing propagators in a constant electromagnetic background field \cite{sdloops,
2L_action}.
   The renormalized two-loop effective action in a constant background field was derived by Ritus \cite{ritus} and later by
   other authors \cite{dit_reu_flie_kors}. The result is a rather complicated double-parameter integral. The extension
   of the "integration by parts" method
  to diagrams in background fields can dramatically simplify the computation
   of the renormalized two-loop effective action \cite{dunnekogan, sdloops, 2L_action}. It has been shown that there is
   a simple diagrammatic interpretation of mass renormalization in the two-loop scalar QED effective action  \cite{2L_action}.
  In the case of a self-dual background field in four dimensions, where
  $F_{\mu\nu}={\widetilde F}_{\mu\nu} \equiv \frac{1}{2} \varepsilon_{\mu\nu\alpha\beta}F_{\alpha\beta}$,
  the field also satisfies $F^2=-f^2 \Eins$, and the whole two-loop Heisenberg-Euler
   effective action acquires a very
   simple form \cite{dunneschubert}, and moreover it can be expressed in terms of one-loop
   quantities \cite{sdloops}. This simplicity reflects the connection between helicity, self-duality and supersymmetry \cite{SD-helicity}.
   The self-duality condition is special to four dimensions; but in this paper we show that the simplicity of the effective action in the
   background field satisfying $F^2=-f^2 \Eins$ persists
in any  even dimension. These results are based on certain recurrence relations, derived in this paper, for two-loop vacuum diagrams in a background field.
 These new relations allow us to obtain the two-loop Heisenberg-Euler effective action completely in terms of one-loop
   integrals in a purely algebraic way, without the need of performing any complicated proper time integrals \cite{dunneschubert}.
An immediate consequence is that we obtain the fully renormalized two-loop effective action
   for any constant background field in two dimensions, since the condition $F^2=-f^2 \Eins$, is satisfied by any
   constant field in two dimensions.

\section{Two-loop effective action}

  Consider a scalar field in Euclidean space, in an electromagnetic background with constant field strength,
  such that the square of the field-strength tensor is proportional to the identity matrix:
\be
     F_{\mu\alpha}F_{\alpha\nu}=-f^2\delta_{\mu\nu}.
\label{background_field}
\ee
Here, $f$ denotes the strength of the field.
In two dimensions this condition is satisfied for an arbitrary constant field, while in four dimensions it is satisfied by a
self-dual constant electromagnetic field.

 The propagator of a scalar field interacting with the background (\ref{background_field}) obeys the Klein-Gordon equation
\ba
(p^2+m^2)G(p)= 1+\frac{\left(ef\right)^2}{4}  \frac{\partial^2
G(p)}{\partial p_\mu \partial p_\mu} \quad .
\label{K-G}
\ea
 The solution of this equation in $d$ dimensions can be written as a proper time integral:
\ba
G(p)= \int_0^{\infty} \!\!\!\! dt  \ \frac{e^{-m^2 t -\frac{p^2}{e f}\tanh(eft)}}{\cosh^{\frac{d}{2}}(eft)}
\label{G}
\ea
 The rotational symmetry of the propagator (\ref{G}) in  Euclidean space eliminates the complicated tensor structure of
 the general two-loop effective action \cite{2L_action} and enables us to express it in a very simple form in terms of
 one-loop diagrams \cite{sdloops, 2L_action}. This rotational symmetry is a direct consequence of the condition
 (\ref{background_field}). We briefly recall the form of the two-loop effective action.

 The background field modifies not only the propagators but also the vertices, such that
 $p_\mu \to p_\mu-i\frac{e}{2}\, F_{\mu\nu}\frac{\partial}{\partial p_\nu}$. Thus, the two-loop bubble diagram is
\ba
\tlscbf \hskip.3cm &=& \frac{e^2}{2} \int \frac{d^dp\, d^dq}{(2\pi)^{2d}}\, \frac{1}{(p-q)^2}\left\{ (p+q)^2G(p)G(q)-
      e^2 f^2 \frac{\partial G(p)}{\partial p_\mu}  \frac{\partial G(q)}{\partial q_\mu} \right\}\,\, .
\label{bfmanip} \ea Here we introduce the notation that the double line denotes the scalar propagator in the background field.

As was shown  \cite{sdloops, 2L_action}, one can express difference between (\ref{bfmanip}) and the corresponding diagram without
the background field by purely algebraic manipulations as
\ba
\Big[\hskip .3cm \mbox{\tlscbf}\hskip .3cm- \hskip
.3cm\mbox{\tlscfr}\hskip .3cm\Big]&=&\frac{e^2}{2}\left( \frac{d-1}{d-3}\right) \Big[\hskip .3cm \mbox{\olscbf}\hskip .3cm-
\hskip .3cm \mbox{\olscfr}\hskip .3cm \Big]^2+
 \left[\hskip .3cm \olmass \hskip .3cm\right]_{p^2=-m^2}
 \Big[\hskip .3cm \mbox{\olscbf}\hskip .3cm - \hskip .3cm \mbox{\olscfr}\hskip .3cm \Big] \nn \\
&-& 2 e^2 m^2 \hskip .3cm \tlscbfph \hskip .3cm
- e^2 \left(\frac{d-2}{d-3}\right) \Big[ \hskip .3cm \olscbf \hskip .3cm \Big]^2 \label{massrenorm}\\
&+& \int \frac{d^dp\, d^dq}{(2\pi)^{2d}}\, \frac{2
e^2}{(p-q)^2}\Big\{(p^2+m^2)+(q^2+m^2)\Big\} G(p) G(q) . \nn \ea
 The dashed line in the first diagram in the second line denotes a free massless scalar, with no contributions from the
vertices. The second term in the first line is the mass renormalization diagram:
\ba
\left[\hskip
.3cm \olmass \hskip .3cm\right]_{p^2=-m^2}=e^2\left(
\frac{d-1}{d-3}\right) \hskip .3cm\olscfr\hskip .3cm \ .
\label{massshift}
\ea
Applying the Klein-Gordon equation (\ref{K-G}) to the last term in (\ref{massrenorm}) and integrating by parts we obtain
\ba
2 e^2 \! \int \frac{d^dp\, d^dq}{(2\pi)^{2d}}\,\frac{(p^2+m^2)+(q^2+m^2) }{(p-q)^2}G(p)G(q)&=&  \label{p^2+q^2} \nn \\
& & \hskip -4cm = \ \ e^4 f^2  \int \frac{d^dp\,
d^dq}{(2\pi)^{2d}}\,\left( \frac{\partial^2}{\partial p_\mu \partial p_\mu}\frac{1}{(p-q)^2}\right) G(p)G(q)  \\
& & \hskip -4cm = 2 e^4 f^2 \Big(4-d\Big) \hskip .3cm \mbox{\tlscbfphvv} \hskip .3cm  \ . \nn
\ea
 The diagram on the RHS contains no contribution from the vertices and the free massless scalar propagator is raised to the second power.
Equation (\ref{p^2+q^2}), together with the recurrence formula (\ref{appendixB n=1}), allows us to write
the two-loop effective action in the self-dual background field in the following form
\ba
 \Big[\hskip .3cm \mbox{\tlscbf}\hskip .3cm- \hskip .3cm\mbox{\tlscfr}\hskip .3cm\Big]&=& \frac{e^2}{2}\left(
\frac{d-1}{d-3}\right) \Big[\hskip .3cm \mbox{\olscbf}\hskip .3cm - \hskip .3cm \mbox{\olscfr}\hskip .3cm \Big]^2
+ \left[\hskip .3cm \olmass \hskip .3cm\right]_{p^2=-m^2}
 \Big[\hskip .3cm \mbox{\olscbf}\hskip .3cm - \hskip .3cm \mbox{\olscfr}\hskip .3cm \Big] \nn \\
 &-& e^4 f^2 \frac{(d-4)(d-2)}{(d-3)} \hskip .3cm \tlscbfphvv \hskip .3cm \ .
\label{2L-action} \ea
All terms in the bare effective action (\ref{2L-action}) are
one-loop except the last one. It has been shown that this diagram
can be reduced in four dimensions to one loop diagrams
\cite{sdloops}.

\section{Analysis of Two-loop Diagrams}
\label{2Ldiagrams}

 In this section we show how the diagrams of the form
\be
\tlscbfphvn \quad = \int \! \frac{d^d p}{(2\pi)^d} \frac{d^d q}{(2\pi)^d} \ \frac{1}{(p-q)^{2n}} G(p) G(q) \label{2L-n}
\ee
can be expressed in terms of one-loop diagrams for any $n$, and that the procedure depends on the dimensionality of space $2N$. When the power of
the free massless scalar propagator is half of the number of dimensions, $n=N$, the finite part of (\ref{2L-n}) is proportional to a one-loop diagram. For any
other power of the free massless scalar propagator, (\ref{2L-n}) can be written as a sum of the previous one and a one-loop diagram.

  The first part comes from the fact that a convergent integral of functions $f(p), g(q)$ and $N$-th power of the free massless scalar propagator
  in $2N$ dimensions can be calculated when the regularization is removed. In $d$ dimensions we can write
\ba
\int \! \frac{d^d p \ d^d q}{(2\pi)^{2d}}  \ \frac{2N-d}{(p-q)^{2N}} f(p) g(q) =
               \frac{1}{2(N-1)} \int \! \frac{d^d p \ d^d q}{(2\pi)^{2d}} \left( \frac{\partial^2}{\partial p_\mu \partial p_\mu}
               \frac{1}{\left( p-q \right)^{2(N-1)}} \right) f(p) g(q) \ . \nn \\
\label{identity2}
\ea
  If the integral converges, we can set $d=2N$, then the derivative of the free massless scalar propagator gives
$ -\frac{4(N-1) \pi^N}{\Gamma(N)} \delta^{2N}(p-q)$ and  the integral reduces to
\ba
 - \frac{1}{2^{2N-1}\pi^N\Gamma(N)}
\int \frac{d^{2N} p}{(2\pi)^{2N}}  f(p) g(p) \ .
\ea
Applying this to the two-loop diagram (\ref{2L-n}) with $n=N$ implies
\ba
(2N-d) \hskip .3cm \tlscbfphvbn \hskip .3cm - g_1 \Big[ \hskip .3cm \olscfr \hskip .3cm \Big]^2 =
                        - \frac{1}{2^{2N-1} \pi^N\Gamma(N)}  \hskip .3cm \olscbfvdva \quad - g_2 \hskip .3cm \olscfr \quad + O(\varepsilon)
\label{2L-1L}
\ea
where $2\varepsilon=2N-d$, and $g_1$ and $g_2$ are known functions of $N,d,m^2$ and $(ef)$. The double-pole divergence of the two-loop vacuum
diagram on the LHS and a single-pole
divergence of the one-loop vacuum diagram on the RHS are subtracted, making both sides of the equation finite.
 Explicit examples of (\ref{2L-1L}) for $N=1,2,3,4$ are given in appendix \ref{appendixC}.

The second important fact is that the two-loop diagram (\ref{2L-n}) with any power $n$ of the free massless scalar propagator can be written as a sum
of  the same diagram with $N$-th power of the free massless scalar propagator and the square of the one-loop diagram:
\ba
\tlscbfphvn \quad = h_1 \quad \tlscbfphvbn \quad + h_2 \Big[ \quad \olscbf \quad \Big]^2 \ .
\label{2L-n=2L-N+1L^2}
\ea
Here, $h_1$ and $h_2$ are simple functions of $n,N,d,m^2$ and $(ef)$.  If $h_1$ is proportional to $2N-d$, we can use (\ref{2L-1L})
and write the diagram (\ref{2L-n}) completely in terms of one-loop diagrams plus a term vanishing as $d\rightarrow 2N$.

  This statement is based on the following recurrence relation proven in appendix \ref{appendixA}
\be
 (ef)^2 n \Big( 2(n+1)-d \Big)^2 \hskip .3cm \tlscbfphvnnn \hskip .3cm - 2 m^2 (2n-d+1) \hskip .3cm \tlscbfphvn \hskip .3cm
 -(n-d+1)\hskip .3cm \tlscbfphvnn \hskip .3cm  =0 \ .
\label{identity}
\ee
 The identity (\ref{identity}) implies that for any $n$ bigger than 1, by successive application of (\ref{identity}),
  the diagram (\ref{2L-n}) can be written as
\ba
 \left( 2n-d \right)^2 \hskip .3cm \tlscbfphvn \hskip .3cm = f_1 \hskip .3cm \tlscbfph \hskip .3cm  + f_2 \Big[ \hskip .3cm \olscbf \hskip .3cm
  \Big]^2 \ ,
\label{2L-n=2L-1_1L}
\ea
 where $f_1$ and $f_2$ are some functions of $n,d,m^2$ and $(ef)$. Relevant examples are given in appendix \ref{appendixB}. If the power of the
free massless scalar propagator is zero, (\ref{2L-n}) turns into the square of the one-loop diagram, producing the second term in (\ref{2L-n=2L-1_1L}).
Because a particular form of (\ref{2L-n=2L-1_1L}) can be found for any $n$, and always contains the two diagrams
$\hskip .3cm \tlscbfph \hskip .3cm$ and $\Big[ \hskip .3cm \olscbf \hskip .3cm \Big]^2$, one can use such equations for $n$ and $N$ to eliminate
$\hskip .3cm \tlscbfph \hskip .3cm$ and obtain (\ref{2L-n=2L-N+1L^2}).

In the case of a vanishing background field the recurrence formula (\ref{identity}) turns into a relation of the vacuum loop diagrams
with the free propagators:
\be
   2 m^2 (2n-d+1) \hskip .3cm \tlscfrphvn \hskip .3cm  + (n-d+1)\hskip .3cm \tlscfrphvnn \hskip .3cm =0 \ .
\label{identity-free}
\ee
This relation gives us a formula equivalent to (\ref{2L-n=2L-1_1L}) for the free propagators:
\ba
\hskip .3cm \tlscfrphvn \hskip .3cm = \frac{1}{m^{2n}} \frac{\Gamma(1-\frac{d}{2}+n) \Gamma(d-1-2n)}{\Gamma(1-\frac{d}{2})\Gamma(d-1-n)}
                  \Big[ \hskip .3cm \olscfr \hskip .3cm \Big]^2 \ .
\label{2L-n-free=1L-free^2}
\ea
By evaluating the one-loop diagram on the RHS of (\ref{2L-n-free=1L-free^2}), one can obtain a special case of Vladimirov's formula \cite{Vladimirov}.

\section{Two-loop Effective Action in $2N$ dimensions}
\label{SD-2L-d}

  The two-loop Heisenberg-Euler effective action (\ref{2L-action}) in the background field (\ref{background_field}) in four dimensions
has been already extensively discussed  \cite{sdloops, 2L_action, dunneschubert}.
In this section we focus on the last two-loop diagram in the effective action (\ref{2L-action}) in different dimensions:
\ba
 &-& e^4 f^2 \frac{(d-4)(d-2)}{(d-3)} \hskip .3cm \tlscbfphvv \hskip .3cm \ .
 \label{charge_ren}
\ea
We show that this diagram can be expressed in terms of single-loop diagrams, using
the identities from the previous section. The particular form of the result depends on the dimensionality of space.

  Splitting the two-loop vacuum diagram (\ref{2L-n}) into one-loop parts
can be done only for its convergent part and only if $n=N$, the power of the photon propagator $n$ is one half of the number of dimensions $2N$.
 Therefore first we use the identity (\ref{identity}) to write (\ref{charge_ren}) as a sum of the second power of the one-loop diagram and
 $ \hskip .3cm \tlscbfphvbn \hskip .3cm$. From this diagram we have to subtract its divergent part that
 can be expressed in terms of  diagrams without the background field (these can always be reduced into one-loop diagrams) and
 the remaining convergent part can be written as a one-loop diagram in the background field. We describe this procedure in 2, 4 and 6 dimensions
 and outline how it can be done in any even dimension.

In two dimensions, $N=1$ and $2 \varepsilon=2-d$. The identity (\ref{identity}) for $n=1$ implies
\ba
- e^4 f^2 \frac{(d-4)(d-2)}{(d-3)} \hskip .3cm \tlscbfphvv \hskip .3cm = e^2  \frac{d-2}{(d-3)(d-4)} \hskip .3cm
                                        \Big\{ 2m^2(d-3) \hskip .3cm \tlscbfph \hskip .3cm
                                     + (d-2) \Big[\hskip .3cm \olscbf \hskip .3cm \Big]^2 \Big\} \ . \nn \\
\label{charge_ren:N=2a}
\ea
Despite the fact that all the diagrams are divergent, they are multiplied by an appropriate power of $d-2$, and thus each term of this equation is finite.
The second term on the RHS is already one-loop. In two dimensions its finite part is the same as for the free propagator:
\begin{displaymath}
e^2  \frac{\left(d-2\right)^2}{(d-3)(d-4)} \Big[\hskip .3cm \olscbf \hskip .3cm \Big]^2 = \frac{e^2}{2} \left( d-2 \right)^2
                                \Big[\hskip .3cm \olscfr \hskip .3cm \Big]^2 +O(\varepsilon) \ ,
\end{displaymath}
 and can be manipulated using the identity (\ref{identity}) for $n=1$ without a background field:
$ \frac{d-2}{d-3} \Big[ \hskip .3cm \olscfr \hskip .3cm \Big]^2 =
- 2 m^2 \hskip .3cm \tlscfrph \hskip .3cm$,
into the same form as the first term on the RHS of (\ref{charge_ren:N=2a}):
\begin{displaymath}
   e^2 m^2(d-2) \hskip .3cm \tlscfrph \hskip .3cm +O(\varepsilon)
\end{displaymath}
Now we can use the equation (\ref{2L-1L:N1}) and an identical equation for a vanishing background field to obtain:
\ba
- e^4 f^2 \frac{(d-4)(d-2)}{(d-3)} \hskip .3cm \tlscbfphvv \hskip .3cm &=&
- e^2 m^2 (d-2)\Big[ \hskip .3cm \tlscbfph \hskip .3cm - \hskip .3cm \tlscfrph \hskip .3cm \Big] + O(\varepsilon) \nn \\ &=&
- \frac{e^2 m^2}{2 \pi} \Big[ \hskip .3cm \olscbfvdva \hskip .3cm - \hskip .3cm \olscfrvdva \hskip .3cm \Big] + O(\varepsilon) \ .
\ea
 In two dimensions (\ref{charge_ren}) does not contain any divergence and after we drop off the mass renormalization from (\ref{2L-action}),
 the renormalized two-loop effective action has the following form
\ba
\Big[\hskip .3cm \mbox{\tlscbf}\hskip .3cm- \hskip .3cm\mbox{\tlscfr}\hskip .3cm\Big]_{ren.} =
- \frac{e^2}{2} \Big[\hskip .3cm \mbox{\olscbf}\hskip .3cm - \hskip .3cm \mbox{\olscfr}\hskip .3cm \Big]^2
 - \frac{e^2 m^2}{2 \pi} \Big[ \hskip .3cm \olscbfvdva \hskip .3cm - \hskip .3cm \olscfrvdva \hskip .3cm \Big] \ .
\label{2L-ren_act-2D}
\ea
This is the finite fully renormalized two-loop effective action in a general constant background field in two dimensions.

In four dimensions, $N=2$ and $2\varepsilon=4-d $, and the part of the effective action containing charge renormalization (\ref{charge_ren})
has already the appropriate form and we do not have to use the identity (\ref{identity}). In this case we can directly
subtract the divergent part and use (\ref{2L-1L:N2}) to obtain
\ba
- e^4 f^2 \frac{(d-4)(d-2)}{(d-3)} \hskip .3cm \tlscbfphvv \hskip .3cm =
- \frac{e^4 f^2}{8 \pi^2} \frac{d-2}{d-3} \Big[ \hskip .3cm \olscbfvdva \hskip .2cm - \hskip .1cm \olscfrvdva \hskip .3cm \Big]
\! - \! e^4 f^2 \frac{(d-4)(d-2)}{(d-3)} \hskip .3cm \tlscfrphvv \hskip .3cm
\! + \! O(\varepsilon).
\nn \\
\ea
The term
\ba
- e^4 f^2 \frac{(d-4)(d-2)}{(d-3)} \hskip .3cm \tlscfrphvv \hskip .3cm =
- \frac{e^4 f^2}{4 m^4} \frac{\left(d-2\right)^2(d-4)}{(d-3)(d-5)}
\Bigg[\hskip .3cm \olscfr \hskip .3cm \Bigg]^2
\ea
corresponds to charge renormalization and so the renormalized two-loop effective action in four dimensions has the form:
\ba
\Big[\hskip .3cm \mbox{\tlscbf}\hskip .3cm- \hskip .3cm\mbox{\tlscfr}\hskip .3cm\Big]_{ren.} =
 \frac{3 e^2}{2} \Big[\hskip .3cm \mbox{\olscbf}\hskip .3cm - \hskip .3cm \mbox{\olscfr}\hskip .3cm \Big]^2
- \frac{e^4 f^2}{4 \pi^2}  \Big[ \hskip .3cm \olscbfvdva \hskip .3cm - \hskip .3cm \olscfrvdva \hskip .3cm \Big] \ ,
\label{2L-ren_act-4D}
\ea
as derived in \cite{sdloops}.

Notice the similarity between this four dimensional result and the previous two dimensional result (\ref{2L-ren_act-2D}).
The fully renormalized effective action in each case, is a linear combination of the same two one-loop terms, but with different coefficients.
Moreover these one-loop terms are closely related. Using the propagator (\ref{G}) in the background field (\ref{background_field}), we find
\ba
\hskip .3cm \olscbf \hskip .3cm - \hskip .3cm \mbox{\olscfr}\hskip .3cm &=&
\frac{\left(ef \right)^{\frac{d}{2}-1}}{\left(4 \pi \right)^{\frac{d}{2}}}
      \int_0^{\infty} \!\!\!\! dt  \ e^{-2\kappa t} \left( \frac{1}{\sinh^{\frac{d}{2}}(t)} - \frac{1}{t^{\frac{d}{2}}} \right)
      \label{1L-1L_free}  \\
\hskip .3cm \olscbfvdva \hskip .3cm - \hskip .3cm \olscfrvdva \hskip .3cm &=&
\frac{\left(ef \right)^{\frac{d}{2}-2}}{\left(4 \pi \right)^{\frac{d}{2}}}
      \int_0^{\infty} \!\!\!\! dt  \ t \ e^{-2\kappa t} \left( \frac{1}{\sinh^{\frac{d}{2}}(t)} - \frac{1}{t^{\frac{d}{2}}} \right) \ ,
      \label{1L2-1L2_free}
\ea
 where $\kappa = \frac{m^2}{2ef}$. From these two equations we see that
\ba
\hskip .3cm \olscbfvdva \hskip .3cm - \hskip .3cm \olscfrvdva \hskip .3cm =
- \frac{1}{2ef} \frac{d}{d\kappa} \Big[\hskip .3cm \mbox{\olscbf}\hskip .3cm - \hskip .3cm \mbox{\olscfr}\hskip .3cm \Big] \ .
\ea
Thus, (\ref{1L2-1L2_free}) is derivative of (\ref{1L-1L_free}) with respect to $\kappa$. In four dimensions, we define
\cite{dunneschubert, sdloops}:
\ba
\xi_{4D}(\kappa) \ \equiv  \   -\frac{\left(4\pi\right)^2}{m^2} \kappa  \Big[  \hskip .3cm \olscbf \hskip .3cm - \hskip .3cm \olscfr \hskip .3cm \Big]_{d=4} =
- \kappa \left(\psi(\kappa) - \ln(\kappa)  + \frac{1}{2\kappa} \right) \ ,
\ea
where $\psi(\kappa)=\frac{d}{d\kappa} \ln \Gamma(\kappa)$ is the Euler digamma function \cite{bateman}. The fully renormalized two-loop
effective action in a self-dual constant background in four dimensions has the form:
\ba
\mathcal S^{(2)}_{d=4} = \alpha \frac{m^4}{\left(4\pi\right)^3} \frac{1}{\kappa^2} \left( \frac{3}{2} \xi^2_{4D}- \xi'_{4D} \right) \ ,
\ea
as was shown in \cite{sdloops, dunneschubert}. Similarly, in two dimensions, we define
\ba
\xi_{2D}(\kappa) \ \equiv  \  4 \pi \Big[ \hskip .3cm \olscbf \hskip .3cm - \hskip .3cm \olscfr \hskip .3cm \Big]_{d=2} = \
- \Big( \psi(\kappa+\half) - \ln(\kappa) \Big)  \ ,
\ea
and the renormalized two-loop effective action in a general constant background field in two dimensions is
\ba
\mathcal S^{(2)}_{d=2} = - \frac{e^2}{32 \pi^2} \Big( \xi^2_{2D} - 4\kappa \ \xi'_{2D} \Big) \ .
\ea

In six dimensions, $N=3$ and  $2\varepsilon=6-d $, and we have to express (\ref{charge_ren}) in terms of $\hskip .3cm \tlscbfphvvv \hskip .3cm$. Its
convergent part then can be reduced to a one-loop diagram. To achieve this, we  use (\ref{appendixB n=1}) and (\ref{appendixB n=2}), and thus for
 (\ref{charge_ren}) we obtain:
\ba
- e^4 f^2 \frac{(d-4)(d-2)}{(d-3)} \hskip .3cm \tlscbfphvv \hskip .3cm &=&
\frac{4 e^4 f^2 m^2}{C_6}\frac{(d-2)\left(d-6\right)^2}{(d-3)} \hskip .3cm \tlscbfphvvv \hskip .3cm
- \frac{e^2}{C_6} \frac{\left(d-2\right)^2}{(d-3)} \Big[\hskip .3cm \olscbf \hskip .3cm \Big]^2 \nn \\
 \text{where} & & C_6= \left(\frac{2m^2}{ef}\right)^2 \left(\frac{5-d}{4-d}\right)+ \left(4-d\right) \ .
\label{charge_ren:N=3}
\ea
The first diagram on the RHS can be written by using (\ref{2L-1L:N3}) as
\ba
(d-6) \hskip .3cm  \tlscbfphvvv \hskip .3cm  =
         \frac{1}{  (4\pi)^3} \Bigg[ \hskip .3cm \olscbfvdva \hskip .3cm - \hskip .3cm \olscfrvdva \hskip .3cm \Bigg]
         + (d-6) \hskip .3cm \tlscfrphvvv \hskip .3cm + O(\varepsilon)
\ea
and (\ref{charge_ren:N=3}) becomes
\ba
- e^4 f^2 \frac{(d-4)(d-2)}{(d-3)} \hskip .3cm \tlscbfphvv \hskip .3cm &=& \!
 \frac{ e^4 f^2 m^2}{2\left(2\pi\right)^3 C_6}\frac{(d-2)(d-6)}{(d-3)}
                       \Bigg[ \hskip .3cm \olscbfvdva \hskip .3cm - \hskip .3cm \olscfrvdva \hskip .3cm \Bigg] \\
&+& \! \frac{4 e^4 f^2 m^2}{C_6}\frac{(d-2)\left(d-6\right)^2}{(d-3)} \hskip .3cm \tlscfrphvvv \hskip .3cm
- \frac{e^2}{C_6} \frac{\left(d-2\right)^2}{(d-3)} \Big[\hskip .3cm \olscbf \hskip .3cm \Big]^2 + O(\varepsilon) \ . \nn
\ea
The difference of the two integrals in the first line is already finite and thus when multiplied by $(d-6)$, it vanishes in the limit
$d\rightarrow 6$. The free diagram in the second term has a double-pole divergence and gives a finite result
when multiplied by $\left(d-6\right)^2$. By virtue of (\ref{2L-n-free=1L-free^2}) it can be reduced to a one-loop diagram.
Thus the two-loop effective action (\ref{2L-action}) in six dimensions has the following form
\ba
 \Big[\hskip .3cm \mbox{\tlscbf}\hskip .3cm- \hskip .3cm\mbox{\tlscfr}\hskip .3cm\Big]&=& \frac{e^2}{2}\left(
\frac{d-1}{d-3}\right) \Big[\hskip .3cm \mbox{\olscbf}\hskip .3cm - \hskip .3cm \mbox{\olscfr}\hskip .3cm \Big]^2
+e^2\left( \frac{d-1}{d-3}\right) \hskip .3cm\olscfr\hskip .3cm
 \Big[\hskip .3cm \mbox{\olscbf}\hskip .3cm - \hskip .3cm \mbox{\olscfr}\hskip .3cm \Big] \nn \\
 &-& \frac{e^2}{C_6} \frac{\left(d-2\right)^2}{(d-3)} \Big[\hskip .3cm \olscbf \hskip .3cm \Big]^2
 + \frac{16 e^4 f^2}{3 m^4 C_6}\left(d-6\right)^2  \Big[\hskip .3cm \olscfr \hskip .3cm \Big]^2 + O(\varepsilon)
\label{2L-action:N=3}
\ea
  The action (\ref{2L-action:N=3}) still contains divergences, but this is because QED in six dimensions is nonrenormalizable.
  Nevertheless, the two-loop effective action in six dimensions is expressed in (\ref{2L-action:N=3}) in terms of one-loop diagrams.

 In this way we could continue to higher  dimensions. In  $2N$ dimensions, $2\varepsilon=2N-d$, the last two-loop term
 (\ref{charge_ren}) of the effective action can be written using the recurrence formula (\ref{identity}) as a sum of
 $\hskip .3cm \tlscbfphvbn \hskip .3cm$ and $\Big[ \hskip .3cm \olscbf \hskip .3cm \Big]^2$. For $N>2$, the first diagram will be always
 multiplied by $\left(2N-d\right)^2$, as can be seen from (\ref{identity}). In such a way there always will be the factor $2N-d$
 in front of $\hskip .3cm \tlscbfphvbn \hskip .3cm$ allowing us to reduce the diagram into a one-loop as described in
 \ref{2Ldiagrams}. Moreover, because the factor in front of it contains the factor $\left(2N-d\right)^2$, the finite part of
 $(2N-d) \hskip .3cm \tlscbfphvbn \hskip .3cm$ will vanish as
 $d\rightarrow 2N$, and its divergent part
 will give a finite contribution to
the effective action. Thus the effective action will always be of the form:
\ba
 \Big[\hskip .3cm \mbox{\tlscbf}\hskip .3cm- \hskip .3cm\mbox{\tlscfr}\hskip .3cm\Big]&=& \frac{e^2}{2}\left(
\frac{d-1}{d-3}\right) \Big[\hskip .3cm \mbox{\olscbf}\hskip .3cm - \hskip .3cm \mbox{\olscfr}\hskip .3cm \Big]^2
+e^2\left( \frac{d-1}{d-3}\right) \hskip .3cm\olscfr\hskip .3cm
 \Big[\hskip .3cm \mbox{\olscbf}\hskip .3cm - \hskip .3cm \mbox{\olscfr}\hskip .3cm \Big] \nn \\
 &-& f_1 \Big[\hskip .3cm \olscbf \hskip .3cm \Big]^2
 + f_2 \left(d-2N\right)^2  \Big[\hskip .3cm \olscfr \hskip .3cm \Big]^2 + O(\varepsilon)
\label{2L-action:N}
\ea
where $f_1$ and $f_2$ are certain known functions of $N,d,m$ and $\left(ef\right)^2$.

\section{Conclusions}
\label{Conclusions}

  In this paper we have further developed the algebraic rules for vacuum diagrams with scalar propagators in  a
  constant electromagnetic field \cite{sdloops, 2L_action}. Such rules are generalizations of the "integration by parts" technique for free propagators
  \cite{Bender, Chetyrkin_Tkachov, Anastasiou, Avdeev, Baikov2, Smirnov_Steinhauser, Schroder, Chetyrkin_Faisst_Sturm_Tentyukov}, a powerful method
  to perform multi-loop calculations. The simplicity of the background field approach opens a way to calculations of higher order loop diagrams in
  a background field.
  We have shown that the reduction of the two-loop Heisenberg-Euler effective action into one-loop diagrams and possible further subtraction
  of divergences can be done in any even number of dimensions. This extends previously obtained results in four dimensions
  \cite{sdloops, 2L_action, ritus, dit_reu_flie_kors, dunneschubert}. We also derived the fully renormalized effective action in a constant
  background field in two dimensions.

\section*{Acknowledgments}

  I thank G. Dunne for discussions, and the US DOE for support through grant DE-FG02-92ER40716.

\appendix

\section{Examples of splitting Two-loop diagram (\ref{2L-n}) into One-loop diagrams.}
\label{appendixC}

In this section we present some examples of (\ref{2L-1L}), the way  to write diagram $\hskip .3cm \tlscbfphvbn \hskip .3cm$ in terms of
one-loop diagrams in a $2N$ dimensional space as was described in \ref{2Ldiagrams}. In the dimensional regularization $2\varepsilon = 2N - d$.

  In two dimensions, $N=1$, the integral on the LHS of (\ref{identity2}), with the propagators $G$ (\ref{G}) in a constant background
  field as functions $f(p)$ and $g(q)$, is convergent. Therefore we obtain
\ba
(d-2)  \hskip .3cm  \tlscbfph \hskip .3cm  =
         \frac{1}{2 \pi}  \hskip .3cm \olscbfvdva \hskip .3cm  + O(\varepsilon)
\label{2L-1L:N1}
\ea
without any need of subtracting divergences. Such subtraction is necessary in higher dimensions. The identity (\ref{2L-1L:N1}) holds also for
zero background field.

 In four dimensions, $N=2$, we have to subtract the divergent part of the two-loop diagram. Then the formula (\ref{2L-1L}) has the following form:
\ba
(d-4) \Bigg[ \hskip .3cm  \tlscbfphvv \hskip .3cm - \hskip .3cm \tlscfrphvv \hskip .3cm \Bigg] =
         \frac{1}{8 \pi^2} \Bigg[ \hskip .3cm \olscbfvdva \hskip .3cm - \hskip .3cm \olscfrvdva \hskip .3cm \Bigg] + O(\varepsilon) \ .
\label{2L-1L:N2}
\ea
 This formula can be brought to a form identical to (\ref{2L-1L}) by using well known identities (\ref{2L-n-free=1L-free^2}) and
 (\ref{1L^n-free=1L-free}).
 The procedure can be repeated in the same way in six dimensions, $N=3$, producing:
\ba
(d-6) \Bigg[ \hskip .3cm  \tlscbfphvvv \hskip .3cm - \hskip .3cm \tlscfrphvvv \hskip .3cm \Bigg] =
         \frac{1}{  (4\pi)^3} \Bigg[ \hskip .3cm \olscbfvdva \hskip .3cm - \hskip .3cm \olscfrvdva \hskip .3cm \Bigg] + O(\varepsilon) \ .
\label{2L-1L:N3}
\ea
In eight dimensions, $N=4$, we have to subtract another divergence to obtain the finite result:
\ba
(d-8) &\Bigg[& \hskip .3cm  \tlscbfphvvvv \hskip .3cm - \hskip .3cm \tlscfrphvvvv \hskip .3cm -
             \left(ef\right)^2 (4-d) \hskip .3cm \tlscfrphvvvvggg \hskip .3cm \Bigg] =
\label{2L-1L:N4} \\
      &&  \hskip 2cm  \frac{1}{3 (4\pi)^4} \Bigg[ \hskip .3cm \olscbfvdva \hskip .3cm - \hskip .3cm \olscfrvdva \hskip .3cm
                   - \left(ef\right)^2 (4-d) \hskip .3cm \olscfrvst \hskip .3cm\Bigg] + O(\varepsilon) \ . \nn
\ea
All the above expressions (\ref{2L-1L:N2}), (\ref{2L-1L:N3}) and (\ref{2L-1L:N4}) can be written in the same form as (\ref{2L-1L}) by using
the equation (\ref{2L-n-free=1L-free^2}) for the loop integrals of the free propagators which can be obtained by integration by parts and
\ba
\hskip .3cm \tlscfrphvvvvggg \hskip .3cm =
\frac{1}{2m^{12}} \frac{\left(6-\frac{d}{2}\right)\left(5-\frac{d}{2}\right)\left(4-\frac{d}{2}\right)\left(3-\frac{d}{2}\right)\left(2-\frac{d}{2}\right)\left(1-\frac{d}{2}\right)}{(d-8)(d-9)(d-10)(d-11)}
\Bigg[\hskip .3cm \olscfr \hskip .3cm \Bigg]^2
\label{2L-3-4-free=1L-free^2}
\ea
together with:
\ba
\hskip .3cm \olscfrvn \hskip .3cm =
\frac{1}{m^{2(n-1)}} \frac{\Gamma(n-\frac{d}{2})}{\Gamma(n) \Gamma(1-\frac{d}{2})}  \hskip .3cm \olscfr \hskip .3cm \ .
\label{1L^n-free=1L-free}
\ea

\section{Proof of the recurrence relation (\ref{identity}).}
\label{appendixA}

A two-loop vacuum diagram of the form (\ref{2L-n}) in the background field (\ref{background_field})
is equal to the following parametric integral
\ba
\tlscbfphvn \hskip .3cm &=& K_n
                 \int_0^\infty dy \ e^{-2\kappa y} \left( \sinh y \right)^{n-d+1}
                 {}_2 F_1(\frac{n}{2}-\frac{d}{4}+\frac{1}{2},\frac{n}{2}-\frac{d}{4}+\frac{1}{2};n-\frac{d}{2}+\frac{3}{2};-\sinh^2 y) \ , \nn \\
           & & \text{where} \qquad   K_n = \frac{1}{2^{2n-d+1}} \frac{(ef)^{d-n-2}}{(4 \pi)^d} \
          \frac{\Gamma(\frac{1}{2})\ \Gamma(n-\frac{d}{2}+1)\ \Gamma(\frac{d}{2}-n)}{\Gamma(\frac{d}{2})\ \Gamma(n-\frac{d}{2}+\frac{3}{2})} \nn \\
           & & \text{and} \qquad  \kappa = \frac{m^2}{2ef} \ .
\label{2L-n:A}
\ea

For $d<n+1$, the value of the following integral is equal to zero
\ba
K_n \int_0^\infty dy \ \frac{d}{dy}\Big[ e^{-2\kappa y} \left( \sinh y \right)^{n-d+1}
                 {}_2 F_1(\frac{n}{2}-\frac{d}{4}+\frac{1}{2},\frac{n}{2}-\frac{d}{4}+\frac{1}{2};n-\frac{d}{2}+\frac{3}{2};-\sinh^2 y) \Big] =0. \nn \\
\ea
By performing the derivative and using the decomposition  $1=\frac{n}{2n-d+1}+\frac{(n-d+1)}{2n-d+1}$ we obtain:
\ba
0 = &-& 2  \kappa \ \hskip .3cm \tlscbfphvn \hskip .3cm  \nn \\
  &+&  \frac{n}{2n-d+1} K_n
         \int_0^\infty \!\!\!\!\! dy \ e^{-2\kappa y} \left( \sinh y \right)^{n-d+1} \nn \\
  & &  \hskip 2cm  \frac{d}{dy} \Big[
          {}_2 F_1(\frac{n}{2}-\frac{d}{4}+\frac{1}{2},\frac{n}{2}-\frac{d}{4}+\frac{1}{2};n-\frac{d}{2}+\frac{3}{2};-\sinh^2 y) \Big]  \\
&+& \frac{n-d+1}{2n-d+1} K_n
                 \int_0^\infty \!\!\!\!\! dy \ e^{-2\kappa y} \frac{1}{\left( \sinh y \right)^n} \nn \\
 & &   \hskip 2cm     \frac{d}{dy} \Big[ \left( \sinh y \right)^{2n-d+1}
                 {}_2 F_1(\frac{n}{2}-\frac{d}{4}+\frac{1}{2},\frac{n}{2}-\frac{d}{4}+\frac{1}{2};n-\frac{d}{2}+\frac{3}{2};-\sinh^2 y) \Big] \nn
\ea
The final step follows from two identities for hypergeometric functions \cite{bateman} :
\ba
&\frac{d}{dy}\left[
{}_2 F_1(\frac{n}{2}-\frac{d}{4}+\frac{1}{2},\frac{n}{2}-\frac{d}{4}+\frac{1}{2};n-\frac{d}{2}+\frac{3}{2};-\sinh^2 y)  \right] = \hskip 4cm  \nn \\
& \hskip 3cm -\frac{1}{4}\frac{\big[ 2(n+1)-d \big]^2}{2n-d+3} \left( \sinh y \right)
{}_2 F_1(\frac{n}{2}-\frac{d}{4}+1,\frac{n}{2}-\frac{d}{4}+1;n-\frac{d}{2}+\frac{5}{2};-\sinh^2 y) \nn \\
&  \frac{d}{dy}\left[ \left( \sinh y \right)^{2n-d+1}
{}_2 F_1(\frac{n}{2}-\frac{d}{4}+\frac{1}{2},\frac{n}{2}-\frac{d}{4}+\frac{1}{2};n-\frac{d}{2}+\frac{3}{2};-\sinh^2 y)  \right] = \qquad \qquad  \\
             &  \hskip 3cm  (2n-d+1)\left( \sinh y \right)^{2n-d}
              {}_2 F_1(\frac{n}{2}-\frac{d}{4},\frac{n}{2}-\frac{d}{4};n-\frac{d}{2}+\frac{1}{2};-\sinh^2 y) \nn
\ea
and the observation that $K_n$ can be written as
\ba
K_n &=& -2ef (2n-d+3) \ K_{n+1} \nn \\
K_n &=& - \frac{1}{2ef(2n-d+1)} \ K_{n-1} \nn
\ea
From the resulting integrals one can quickly recover the two-loop integrals (\ref{2L-n:A}) with $(n+1)$-th and $(n-1)$-th power of
the free massless scalar
propagator and obtain the formula (\ref{identity}). By analytic continuation we can extend region of validity of (\ref{identity}) beyond
$d<n+1$.

\section{Examples of the recurrence formula (\ref{identity}).}
\label{appendixB}

  In this section we give some examples of the relation (\ref{identity}) for $n=1,2,3$ and also the way (\ref{2L-n=2L-1_1L}) of writing
  the two-loop diagram $\hskip .3cm \tlscbfphvn \hskip .3cm$ in terms of the diagrams $\hskip .3cm \tlscbfph \hskip .3cm$ and
  $\hskip .3cm \olscbf \hskip .3cm$.
  For $n=1$, the identity (\ref{identity}) acquires the following form:
\ba
 \left( ef \right)^2 \left( d-4 \right)^2 \hskip .3cm \tlscbfphvv \hskip .3cm -
 2m^2(3-d)\hskip .3cm \tlscbfph \hskip .3cm
-(2-d)\Big[ \hskip .3cm \olscbf \hskip .3cm \Big]^2 = 0
\label{appendixB n=1}
\ea
   For $n=2$, the identity (\ref{identity}) has the following form:
\ba
2 \left( ef \right)^2 \left( d-6 \right)^2  \hskip .3cm \tlscbfphvvv \hskip .3cm -
 2m^2(5-d)\hskip .3cm \tlscbfphvv \hskip .3cm
- (3-d) \hskip .3cm \tlscbfph \hskip .3cm  = 0
\label{appendixB n=2}
\ea
 We can use (\ref{appendixB n=1}) to write (\ref{appendixB n=2}) as
\ba
\hskip .3cm \tlscbfphvvv \hskip .3cm =
\frac{1}{\left( ef \right)^2} \frac{1}{2 \left( d-6 \right)^2} & \Bigg\{& \Bigg[ \left( \frac{2m^2}{ef} \right)^2
 \frac{(5-d)(3-d)}{\left(d-4\right)^2} +(3-d) \Bigg] \hskip .3cm \tlscbfph \hskip .3cm \label{appendixB n=2b} \\
&+& \frac{2m^2}{\left(ef\right)^2} \frac{(5-d)(2-d)}{\left(d-4\right)^2} \Bigg[ \hskip .3cm \olscbf \hskip .3cm \Bigg]^2  \Bigg\} \ . \nn
\ea
 For $n=3$, the identity (\ref{identity}) has the following form:
\ba
3 \left( ef \right)^2  \left( d-8 \right)^2 \hskip .3cm \tlscbfphvvvv \hskip .3cm - 2m^2(7-d)\hskip .3cm \tlscbfphvvv \hskip .3cm
- (4-d) \hskip .3cm \tlscbfphvv \hskip .3cm  = 0
\label{appendixB n=3}
\ea
 Now we use (\ref{appendixB n=1}) and (\ref{appendixB n=2b}) to write (\ref{appendixB n=3}) as
\ba
\hskip .3cm \tlscbfphvvvv \hskip .3cm &=&
\frac{1}{\left( ef \right)^2} \frac{1}{3 \left( d-8 \right)^2} \Bigg\{ \frac{1}{2ef} \\
& & \hskip-1cm \Bigg[  \left( \frac{2m^2}{ef} \right)^3
 \frac{(7-d)(5-d)(3-d)}{\left(d-6\right)^2 \left(d-4\right)^2}
 + \frac{2m^2}{ef} \frac{(7-d)(4-d)(3-d)+ 2 \left(6-d\right)^2 (3-d)}{\left(d-6\right)^2 \left(4-d\right)}
                                             \Bigg] \hskip .3cm \tlscbfph \hskip .3cm \nn \label{appendixB n=3b} \\
&+&\frac{1}{\left(ef\right)^2} \Bigg[ \left( \frac{2m^2}{ef}\right)^2 \frac{(7-d)(5-d)(2-d)}{2\left(d-6\right)^2 \left(d-4\right)^2}
+ \frac{2-d}{4-d} \Bigg]
\Bigg[ \hskip .3cm \olscbf \hskip .3cm \Bigg]^2  \Bigg\}  \nn
\ea
In this way we can continue and express the two-loop diagram of the form (\ref{2L-n}) for any $n$ as a linear combination of the diagram
$\hskip .3cm \tlscbfph \hskip .3cm$ and the square of the diagram $\hskip .3cm \olscbf \hskip .3cm$.


\begin{thebibliography}{01234}


\bibitem{Smirnov}
  V.~A.~Smirnov,
  ``Evaluating Feynman Integrals,''
  Springer Tracts Mod.\ Phys.\  {\bf 211}, 1 (2004).

\bibitem{Steinhauser}
  M.~Steinhauser,
  ``Results And Techniques Of Multi-Loop Calculations,''
  Phys.\ Rept.\  {\bf 364}, 247 (2002)
  [arXiv:hep-ph/0201075].

\bibitem{Baikov}
  P.~A.~Baikov,
  ``Advanced methods of multi-loop integrals calculations: Status and perspectives,''
  Nucl.\ Phys.\ Proc.\ Suppl.\  {\bf 116}, 378 (2003);

\bibitem{Bern}
  Z.~Bern,
  ``Recent Progress In Perturbative Quantum Field Theory.''
  Nucl.\ Phys.\ Proc.\ Suppl.\  {\bf 117}, 260 (2003)
  [arXiv:hep-ph/0212406].

\bibitem{Tarasov}
  O.~V.~Tarasov,
   ``Reduction of Feynman graph amplitudes to a minimal set of basic integrals,''
  Acta Phys.\ Polon.\ B {\bf 29}, 2655 (1998)
  [arXiv:hep-ph/9812250].
  ``Connection between Feynman integrals having different values of the space-time dimension,''
  Phys.\ Rev.\ D {\bf 54}, 6479 (1996)
  [arXiv:hep-th/9606018].

\bibitem{Kotikov}
  A.~V.~Kotikov,
   ``Differential equations method: New technique for massive Feynman diagrams calculation,''
  Phys.\ Lett.\ B {\bf 254}, 158 (1991);
   ``Some Methods For The Evaluation Of Complicated Feynman Integrals,''
  arXiv:hep-ph/0112347.

\bibitem{Laporta}
  S.~Laporta,
   ``High-precision calculation of multi-loop Feynman integrals by  difference equations,''
  Int.\ J.\ Mod.\ Phys.\ A {\bf 15}, 5087 (2000)
  [arXiv:hep-ph/0102033].

\bibitem{Broadhurst}
  D.~J.~Broadhurst,
  ``The master two loop diagram with masses,''
  Z.\ Phys.\ C {\bf 47}, 115 (1990).
  ``Three loop on-shell charge renormalization without integration: Lambda-MS (QED) to four loops,''
  Z.\ Phys.\ C {\bf 54}, 599 (1992).

\bibitem{Vladimirov}
  A.~A.~Vladimirov,
   ``Method For Computing Renormalization Group Functions In Dimensional Renormalization Scheme,''
  Theor.\ Math.\ Phys.\  {\bf 43}, 417 (1980)
  [Teor.\ Mat.\ Fiz.\  {\bf 43}, 210 (1980)].


\bibitem{Bender}
  C.~M.~Bender, R.~W.~Keener and R.~E.~Zippel,
   ``New Approach To The Calculation Of F(1) (Alpha) In Massless Quantum Electrodynamics,''
  Phys.\ Rev.\ D {\bf 15}, 1572 (1977).

\bibitem{Chetyrkin_Tkachov}
  K.~G.~Chetyrkin and F.~V.~Tkachov,
   ``Integration By Parts: The Algorithm To Calculate Beta Functions In 4 Loops,''
  Nucl.\ Phys.\ B {\bf 192}, 159 (1981);

\bibitem{Anastasiou}
  C.~Anastasiou and A.~Lazopoulos,
  ``Automatic integral reduction for higher order perturbative  calculations,''
  JHEP {\bf 0407}, 046 (2004)
  [arXiv:hep-ph/0404258].


\bibitem{Avdeev}
  L.~V.~Avdeev,
   ``Recurrence Relations For Three-Loop Prototypes Of Bubble Diagrams With A Mass,''
  Comput.\ Phys.\ Commun.\  {\bf 98}, 15 (1996)
  [arXiv:hep-ph/9512442].
  L.~V.~Avdeev, J.~Fleischer, M.~Y.~Kalmykov and M.~N.~Tentyukov,
  ``Towards automatic analytic evaluation of diagrams with masses,''
  Comput.\ Phys.\ Commun.\  {\bf 107}, 155 (1997)
  [arXiv:hep-ph/9710222].

\bibitem{Baikov2}
  P.~A.~Baikov,
  ``Explicit solutions of the 3--loop vacuum integral recurrence relations,''
  Phys.\ Lett.\ B {\bf 385}, 404 (1996)
  [arXiv:hep-ph/9603267];
  P.~A.~Baikov and M.~Steinhauser,
  ``Three-loop vacuum integrals in FORM and REDUCE,''
  Comput.\ Phys.\ Commun.\  {\bf 115}, 161 (1998)
  [arXiv:hep-ph/9802429].

\bibitem{Smirnov_Steinhauser}
  V.~A.~Smirnov and M.~Steinhauser,
  ``Solving recurrence relations for multi-loop Feynman integrals,''
  Nucl.\ Phys.\ B {\bf 672}, 199 (2003)
  [arXiv:hep-ph/0307088].

\bibitem{Schroder}
  Y.~Schroder,
  ``Automatic Reduction Of Four-Loop Bubbles,''
  Nucl.\ Phys.\ Proc.\ Suppl.\  {\bf 116}, 402 (2003)
  [arXiv:hep-ph/0211288].

\bibitem{Chetyrkin_Faisst_Sturm_Tentyukov}
  K.~G.~Chetyrkin, M.~Faisst, C.~Sturm and M.~Tentyukov,
  ``e-finite basis of master integrals for the integration-by-parts method,''
  Nucl.\ Phys.\ B {\bf 742}, 208 (2006)
  [arXiv:hep-ph/0601165].


\bibitem{dunnekogan}
  G.~V.~Dunne,
 ``Heisenberg-Euler effective Lagrangians: Basics and extensions,''  in Ian Kogan Memorial Collection,
 {\it From Fields to Strings: Circumnavigating Theoretical Physics}, Vol. I, M.~Shifman (ed.) et al, (World Scientific, 2005)  [arXiv:hep-th/0406216].

\bibitem{sdloops}
  G.~V.~Dunne,
     ``Two-loop diagrammatics in a self-dual background,''
     JHEP {\bf 0402}, 013 (2004)  [arXiv:hep-th/0311167].

\bibitem{2L_action}
  G.~V.~Dunne and M.~Krasnansky,
   ``'Background field integration-by-parts' and the connection between one-loop and two-loop Heisenberg-Euler effective actions,''
  JHEP {\bf 0604}, 020 (2006)
  [arXiv:hep-th/0602216].


\bibitem{ritus}
 V.~I.~Ritus,
   ``Lagrangian Of An Intensive Electromagnetic Field And Quantum Electrodynamics At Short Distances,''
              Sov.\ Phys.\ JETP {\bf 42}, 774 (1975)  [Pisma Zh.\ Eksp.\ Teor.\ Fiz.\  {\bf 69}, 1517 (1975)];
   ``On The Relation Between The Quantum Electrodynamics Of An Intense Field And The Quantum Electrodynamics At Small Distances.,''
           Zh.\ Eksp.\ Teor.\ Fiz.\  {\bf 73}, 807 (1977);
``The Lagrangian Function of an Intense Electromagnetic Field'', in {\it Proc. Lebedev Phys. Inst.} Vol. {\bf 168}, {\it
              Issues in Intense-field Quantum Electrodynamics}, V. I. Ginzburg, ed., (Nova Science Pub., NY 1987).

\bibitem{dit_reu_flie_kors}
  W.~Dittrich and M.~Reuter,
   ``Effective Lagrangians In Quantum Electrodynamics,''
                               Lect.\ Notes Phys.\  {\bf 220}, 1 (1985);
  M.~Reuter, M.~G.~Schmidt and C.~Schubert,
     ``Constant external fields in gauge theory and the spin 0, 1/2, 1 path  integrals,''
                     Annals Phys.\  {\bf 259}, 313 (1997)  [arXiv:hep-th/9610191];
  D.~Fliegner, M.~Reuter, M.~G.~Schmidt and C.~Schubert,
   ``Two-loop Euler-Heisenberg Lagrangian in dimensional regularization,''
                     Theor.\ Math.\ Phys.\  {\bf 113}, 1442 (1997)  [Teor.\ Mat.\ Fiz.\  {\bf 113}, 289 (1997)]
                     [arXiv:hep-th/9704194];
    B.~Kors and M.~G.~Schmidt,
   ``The effective two-loop Euler-Heisenberg action for scalar and spinor  QED in a general constant background field,''
                    Eur.\ Phys.\ J.\ C {\bf 6}, 175 (1999)  [arXiv:hep-th/9803144].

\bibitem{dunneschubert}
  G.~V.~Dunne and C.~Schubert,
       ``Closed-form two-loop Euler-Heisenberg Lagrangian in a self-dual background,''
  Phys.\ Lett.\ B {\bf 526}, 55 (2002)
  [arXiv:hep-th/0111134];
   ``Two-loop self-dual Euler-Heisenberg Lagrangians. I: Real part and  helicity  amplitudes,''
  JHEP {\bf 0208}, 053 (2002)
  [arXiv:hep-th/0205004];
   ``Two-loop self-dual Euler-Heisenberg Lagrangians. II: Imaginary part  and  Borel analysis,''
  JHEP {\bf 0206}, 042 (2002)
  [arXiv:hep-th/0205005].

\bibitem{SD-helicity}
    M.~J.~Duff and C.~J.~Isham,
   ``Selfduality, Helicity, And Supersymmetry: The Scattering Of Light By Light,''
    Phys.\ Lett.\ B {\bf 86}, 157 (1979);
      A.~D'Adda and P.~Di Vecchia,
  ``Supersymmetry and instantons,
  Phys.\ Lett.\ B {\bf 73}, 162 (1978);
  I.~Bialynicki-Birula, E.~T.~Newman, J.~Porter, J.~Winicour, B.~Lukacs, Z.~Perjes and A.~Sebestyen,
   ``A Note On Helicity,''
   J.\ Math.\ Phys.\  {\bf 22}, 2530 (1981).


\bibitem{bateman} A. Erd\'elyi (ed.), {\it Higher Transcendental Functions, Vol. I}, (Kreiger, Florida, 1981).


\end{thebibliography}
\end{document}